\documentclass[sigconf]{acmart}

\AtBeginDocument{%
  }

\setcopyright{acmlicensed}
\copyrightyear{2018}
\acmYear{2018}
\acmDOI{XXXXXXX.XXXXXXX}

\acmConference[Conference acronym 'XX]{Make sure to enter the correct
  conference title from your rights confirmation emai}{June 03--05,
  2018}{Woodstock, NY}
\acmISBN{978-1-4503-XXXX-X/18/06}

\usepackage{graphicx}
\usepackage{tikz}
\usepackage{geometry}
\usepackage{xcolor}
\usepackage[T1]{fontenc}
\usepackage{tabularx}
\usepackage{longtable}
\usepackage{titlesec}
\usepackage{fontawesome}
\usepackage{xcolor}
\usepackage{array}
\usepackage{siunitx}
\usepackage{booktabs}
\usepackage{caption}

\begin{document}

\title{Tracking the 2024 US Presidential Election Chatter on TikTok: A Public Multimodal Dataset}


\author{%
  Gabriela Pinto\textsuperscript{1}, 
  Charles Bickham\textsuperscript{1}, 
  Tanishq Salkar\textsuperscript{1}, 
  Joyston Menezes\textsuperscript{1}, 
  Luca Luceri\textsuperscript{1}, 
  Emilio Ferrara\textsuperscript{1}%
}

\affiliation{%
  \institution{University of Southern California}
  \city{Los Angeles}
  \state{California}
  \country{USA}
}
\email{gpinto@usc.edu, cbickham@usc.edu, salkar@usc.edu, jmenezes@usc.edu, lluceri@usc.edu, emiliofe@usc.edu}

\renewcommand{\shortauthors}{Pinto et al.}

\begin{abstract}
This paper presents the TikTok 2024 U.S. Presidential Election Dataset, a large-scale, resource designed to advance research into political communication and social media dynamics. The dataset comprises 3.14 million videos published on TikTok between November 1, 2023, and October 16, 2024, encompassing video ids and transcripts. Data collection was conducted using the TikTok Research API with a comprehensive set of election-related keywords and hashtags, supplemented by third-party tools to address API limitations and expand content coverage, enabling analysis of hashtag co-occurrence networks that reveal politically aligned hashtags based on ideological affiliations, the evolution of top hashtags over time, and summary statistics that highlight the dataset's scale and richness. This dataset offers insights into TikTok’s role in shaping electoral discourse by providing a multimodal view of election-related content. It enables researchers to explore critical topics such as coordinated messaging, misinformation spread, audience engagement, and linguistic trends. The TikTok 2024 U.S. Presidential Election Dataset is publicly available and aims to contribute to the broader understanding of social media's impact on democracy and public opinion.

\end{abstract}




\maketitle

\section{Introduction}
Social media has profoundly reshaped electoral politics, emerging as a critical medium for disseminating election-related information and influencing public opinion. This shift has garnered significant attention from researchers \cite{jungherr2016twitter, kratzke2017btw17, deb2019perils, luceri2019evolution, abilov2021voterfraud2020}. Although platforms like Twitter have long been a cornerstone in the study of global geopolitical events \cite{davis2016osome, chen2022election2020, chen2023tweets}, TikTok offers a distinctive perspective on political discourse. Its short-form video format, combined with interactive features like comments, hashtags, and trends, enables rapid content dissemination and highly visual engagement, making it uniquely suited for studying political communication. With an estimated global user base exceeding one billion,\footnote{\url{https://whatsthebigdata.com/tiktok-statistics/}} TikTok has quickly become a platform where political campaigns engage and inform audiences, particularly adolescents \cite{montag2021psychology}.

The 2024 U.S. Presidential Election highlighted TikTok’s growing influence in electoral politics, as major candidates actively engaged with the platform to connect with voters. For instance, Donald Trump joined TikTok in June 2024, rapidly gaining over 6 million followers,\footnote{\url{https://cnn.com/2024/06/02/politics/donald-trump-joins-tiktok/index.html}} while Joe Biden actively posted over 200 videos starting in February 2024, amassing more than 373,000 followers.\footnote{\url{https://cnn.com/2024/06/12/tech/tiktok-pew-research-politics-x/index.html}} These examples illustrate TikTok’s potential as both a campaign tool and a rich data source for understanding electoral sentiment, political narratives, and voter outreach strategies. Its unique format and audience demographic make TikTok an important platform for studying the evolving nature of political communication in the digital age.

This paper introduces the TikTok 2024 U.S. Presidential Election Dataset, a collection of over 3 million videos that includes metadata and textual transcripts. The dataset captures the dynamics of TikTok’s election-related discourse, enabling researchers to explore key questions such as identifying coordinated messaging among political actors, detecting the spread of misinformation, examining audience engagement patterns, and analyzing discourse across linguistic and demographic groups.

By leveraging this dataset, researchers can gain a deeper understanding of how political communication evolves in digital spaces, particularly on TikTok’s unique platform. It highlights TikTok’s growing influence in shaping public opinion, especially among younger voters—a pivotal demographic in modern election. Ultimately, this dataset seeks to contribute to broader discussions about the role of digital platforms in electoral processes and the implications for digital democracy in the age of social media. 



\section{Method of Data Collection}
\subsection*{TikTok API and Data Collection}

The TikTok Research API\footnote{\url{https://developers.tiktok.com/doc/research-api-specs-query-videos/}} provides access to both account-related and content-related data on TikTok. Account data includes user profiles, follower lists, liked videos, and reposts, while content data encompasses video descriptions, hashtags, comments, subtitles, and engagement metrics such as likes, shares, and comments. Accessing the API requires a bearer token, generated using client and secret keys, which must accompany API requests. After exploring third-party scrapers, the Official Research API proved to be the most reliable method for metadata collection.

\subsubsection*{Video Metadata Collection.}
We used the TikTok Research API to collect metadata for videos, leveraging parameterized queries to target content by regions, topics, languages, or publication dates. Key parameters included the date range, defining the start and end dates for data collection, and fields, which specified video ID, description, creation time, region, engagement metrics (likes, views, comments, shares), music ID, hashtags, and username. Queries were restricted to 100 records per call, and advanced filters, using logical operators (e.g., AND/OR), allowed precise targeting of content based on specific hashtags and regions. For instance, videos tagged with \#election2024 in the U.S. were queried using region\_code=``US'' and hashtag=``\#election2024''.

The metadata collection process was structured around phases defined by significant political events, with the project divided into several phases spanning specified date ranges. Phase 1 spans from November 1, 2023, to January 15, 2024. Phase 2 continues from January 16, 2024, to February 7, 2024, followed by Phase 3, which occurs between February 8, 2024, and February 27, 2024. Phase 4 is scheduled from February 28, 2024, to March 5, 2024. Subsequently, Phase 5 runs from March 6, 2024, to July 14, 2024, and Phase 6 takes place between July 15, 2024, and July 20, 2024. Phase 7 follows shortly after, from July 20, 2024, to August 5, 2024. Phase 8 is set for the period of August 6, 2024, to September 9, 2024, and Phase 9 will occur between September 10, 2024, and September 15, 2024. Phase 10 is scheduled from September 15, 2024, to November 10, 2024, and finally, Phase 11 begins on November 11, 2024, and continues to the present.

\subsection*{Video Downloads and Transcript Generation}
To complement the collected metadata, we used PykTok\footnote{\url{https://github.com/dfreelon/pyktok}}, a Python library designed to interact with TikTok’s platform, to download the corresponding video content. After downloading, we utilized OpenAI Whisper \footnote{\url{https://github.com/openai/whisper}} to generate transcripts for each video. To ensure accuracy, we randomly selected a subset of 100 videos and verified the correctness of their generated transcripts.

\textbf{Note:} The data collection process complies fully with TikTok's Terms of Service, ensuring adherence to ethical standards.

\section{Exploratory Analysis}
Using the collected data, we performed an exploratory analysis to identify key patterns and trends in TikTok's election-related discourse. This analysis offers an initial glimpse into how users engaged with and discussed major political events and candidates during the 2024 U.S. Presidential Election.

\vspace{-.3cm}

\subsection*{Summary of Data}
Table~\ref{tab:sumStats} summarizes the statistics for the current dataset. TikTok-generated transcripts account for only 15.4\% of the total videos, reflecting a limitation of the TikTok Research API. To enrich the dataset with more linguistic data, we incorporate Whisper-generated transcripts to our dataset and related GitHub repository.


\begin{table}[t]
\footnotesize
\begin{tabular}{|l|l|}
    \hline
    Number  of videos & 3,144,836 \\ \hline  
    Number  of transcripts & 485,343 \\ \hline 
    Number  of comments & 161,892,802 \\ \hline
    Number  of views   & 18,010,334,100   \\ \hline
    Number  of likes   & 2,042,078,329   \\ \hline
    Number  of shares   & 119,934,718   \\ \hline
    Number  of unique users & 693,826 \\ \hline
\end{tabular}
\caption{Summary statistics of the dataset.}
\label{tab:sumStats}
\end{table}

\subsection*{Number of videos over time}
\begin{figure}[t]
\centering
\includegraphics[clip, trim=3 3 0 20, width=\columnwidth]{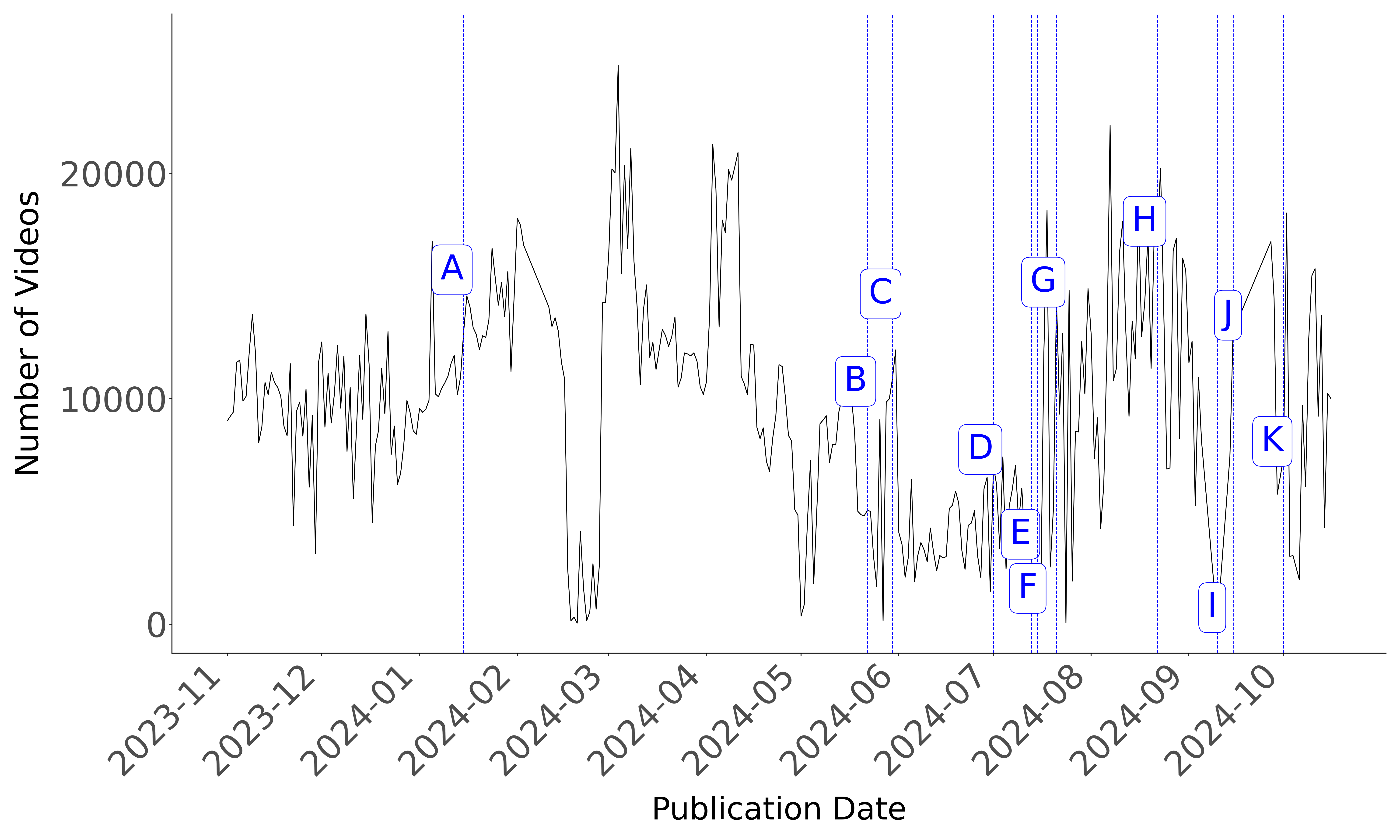}
\centering \footnotesize
\begin{tabularx} {\columnwidth}{|l|l|X|} %
\hline
\textbf{\#} & \textbf{Date} & \textbf{Description} \\ 
\hline

A & 01-15-2024 & Trump wins the Iowa Caucus \\ 
\hline

B & 05-22-2024 & Biden announces a student loans debt cancellation of \$7.7 billion for 160,000 people \\
\hline 
C & 05-30-2924 & Trump is found guilty of 34 counts of falsifying business records in his New York hush money trial \\
\hline

D & 07-01-2024 & In Trump v. United States, the Supreme Court rules 6-3 that former presidents have absolute immunity for core constitutional acts, presumed immunity for other official acts, and none for unofficial acts. \\
\hline
E & 07-13-2024 & Trump is shot in an assassination attempt at a campaign rally in Butler, Pennsylvania. \\
\hline
F & 07-15-2024 & Trump announced his running-mate, Senator JD Vance from Ohio, at the Republican National Convention \\
\hline 
G & 07-21-2024 & President Biden announces his withdrawal from the 2024 Presidential election and Vice President Kamala Harris launches her campaign for president\\
\hline 
H & 08-22-2024 & Kamala Harris and Tim Waltz are officially nominated as the party's choice for President and Vice President\\
\hline
I & 09-10-2024 & Trump and Harris debate for the first time at the National Constitution Center in Philadelphia \\
\hline 
J & 09-15-2024 & An assassination attempt against Trump takes place at West Palm Beach Golf Club\\
\hline
K & 10-01-2024 & The vice presidents, Tim Waltz and JD Vance, debate in New York City \\
\hline 
\end{tabularx}
\caption{Timeline of events and volume of TikTok posts.}
\label{fig:videostime}
\vspace{.5cm}
\end{figure}

This dataset includes video metadata collected for posts published between November 1, 2023, and October 16, 2024. Figure~\ref{fig:videostime} shows the number of videos collected by publication date, along with key political events during the election cycle and the corresponding number of posts on those dates. The data collection process will conclude after the Inauguration Day in January 2025.

\setlength{\belowcaptionskip}{-0.7cm}

\vspace{-.2cm}  

\subsection*{Top Hashtags}
Table~\ref{tab:tophastags} presents the 20 most frequent hashtags for each video. Each video collected through the TikTok Research API provides a list of hashtags labeled ``hashtag\_names'' Overall, the hashtags reflect that the videos labeled by the creator were related to Donald Trump, President Joe Biden, and the right-leaning ideology through keywords such as ``maga'' (``Make America Great Again'') and ``republican''

\begin{table}[t]
\centering\footnotesize
\begin{tabular}{ll}
\toprule
\textbf{Tag} & \textbf{Count} \\
\midrule
trump        & 774,506 \\
trump2024    & 465,388 \\
maga         & 349,150 \\
biden        & 338,506 \\
republican   & 263,289 \\
donaldtrump  & 261,756 \\
usa          & 234,190 \\
politics     & 204,580 \\
democrat     & 181,544 \\
america      & 175,687 \\
kamalaharris & 157,783 \\
joebiden     & 150,204 \\
news         & 146,906 \\
voteblue     & 144,656 \\
election     & 141,000 \\
democrats    & 135,433 \\
conservative & 134,280 \\
vote         & 126,958 \\
kamala       & 115,919 \\
election2024 & 109,770 \\
\bottomrule
\end{tabular}
\vspace{4em} 
\caption{Top 20 Most Frequently Occurred Hashtags.}
\label{tab:tophastags}
\end{table}


\subsection*{Language Detection}
We used LangDetect\footnote{\url{https://pypi.org/project/langdetect/}} to identify the languages of the transcripts generated by the TikTok Research API. In our current dataset, 460,735 transcripts have been classified as English-based content, making it the dominant language. The second most common language is Spanish, with 15,621 transcripts. In future work, we plan to analyze Spanish-based content further to provide a more comprehensive view of the discourse surrounding the election cycle.

\subsection*{Hashtag Co-Occurrence Graph}

\begin{figure}[t]
  \centering
\includegraphics[width=\columnwidth]{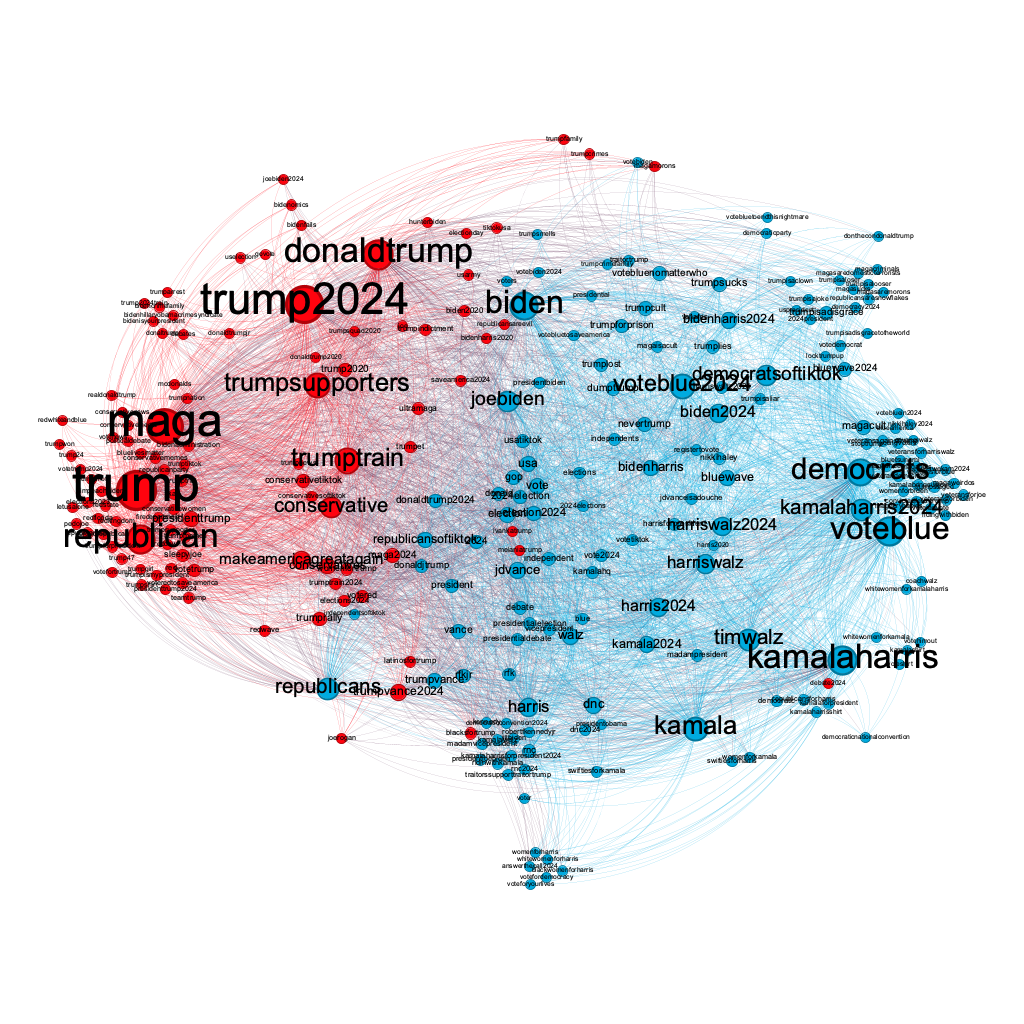}
  \caption{Hashtag Co-Occurrence Graph}
  \label{fig:hashtagcooccur}
  \vspace{5mm}
\end{figure}

Figure \ref{fig:hashtagcooccur} shows a hashtag co-occurrence graph of TikTok content related to the 2024 U.S. Election. We selected the top 35 left-leaning and right-leaning hashtags, including additional hashtags that co-occurred with them. Hashtag pairs appearing together more than 500 times were retained, and the network was visualized in Gephi\footnote{\url{https://gephi.org/}} using the modularity class algorithm, revealing a political divide.

The Right-Leaning cluster (Red) includes interconnected hashtags like \#trumpsupporters, \#trumptrain, \#trump, \#donaldtrump, and \#maga, reflecting strong support for Trump and Republican themes. The Left-Leaning cluster (Blue) features hashtags such as \#kamalaharris, \#voteblue, \#voteblue2024, \#bluewave, and \#biden, emphasizing Democratic themes and voter mobilization.

\subsection*{Hashtags Over Time}
\begin{figure*}[t]
\includegraphics[width=1.5\columnwidth]{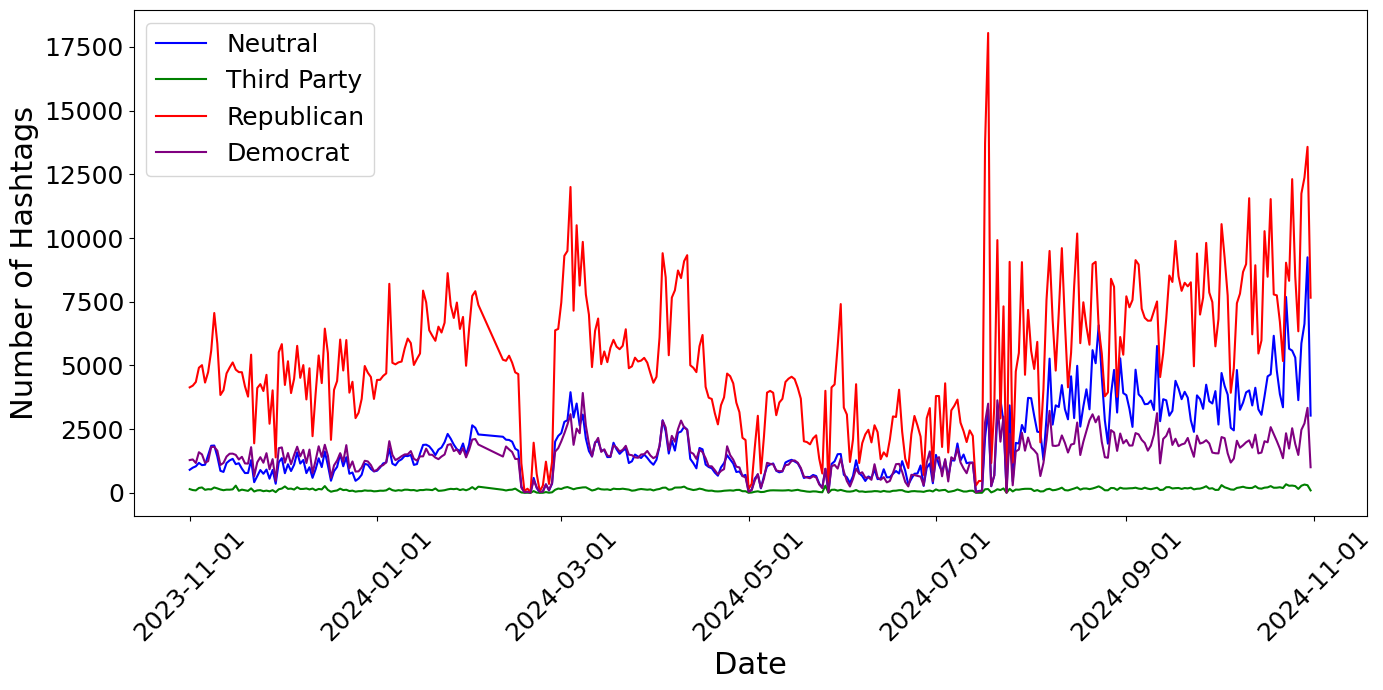}
\caption{\small Hashtag usage trends by category from November 2023 to October 2024, showing peak activity for Republicans, Democrats, Neutral, and Third-Party hashtags over time.}
\label{fig:hashtagsovertime}
\vspace{.5cm}
\end{figure*}

Figure \ref{fig:hashtagsovertime} reveals significant trends across political and neutral categories. Republicans accounted for the majority of hashtags, totaling 1,882,321, with a notable peak of 18,045 hashtags on July 18, 2024. Neutral hashtags were the second most common, reaching a peak of 9,243 on October 30, 2024, and accumulating 696,917 overall. Democrats followed with a total of 524,664 hashtags, peaking at 3,918 on March 8, 2024. Third-party hashtags were minimal at 44,980 total, peaking at 336 on October 23, 2024. The ``Other'' category saw an extraordinary spike on March 4, 2024, with 243,128 hashtags. This data underscores the dynamic engagement trends across political affiliations and neutral discussions, with Republicans dominating both overall volume and peak activity periods.



\subsection*{Topic Modeling}
To analyze recurring themes related to the 2024 US Presidential election, we conducted topic modeling and summarization of TikTok transcripts from April to November 2024. Using BERTopic for topic extraction and the LLaMA 3.1 8B-Instruct model for summarization, we examined weekly discussions, filling in missing data with Whisper. Immigration emerged as a key theme, with bilingual debates on unresolved issues at the US-Mexico border and dissatisfaction with government responses. Social issues such as gun violence, systemic corruption, reproductive rights, and economic struggles were also prominent, reflecting widespread user concerns. Religion featured in discussions about Catholic and Muslim values, religious texts, and their implications for politics. 

Conspiracy theories were prevalent, including speculative discussions about security breaches, assassination attempts, and media bias, with concerns over misinformation frequently highlighted. International conflicts, such as the Israel-Palestine conflict and the Ukraine war, also drew attention, particularly regarding the US role in these crises. Political figures dominated many conversations, with Donald Trump’s legal challenges, policies, and reelection prospects sparking mixed sentiments, while Joe Biden faced criticism over immigration laws, inflation, and rumors of withdrawal from the race. Kamala Harris garnered discussions about her fundraising efforts, policies, and potential to become the first female US president. This analysis provides a comprehensive view of the diverse political and social narratives shaping TikTok conversations during the 2024 election cycle.


\section{Conclusions}
The TikTok 2024 U.S. Presidential Election Dataset offers a large-scale resource for advancing research into political communication and social media dynamics. Spanning 3.14 million videos collected using the TikTok Research API and third-party tools, it provides a comprehensive view of election-related content from November 1, 2023, to October 16, 2024. This dataset enables the analysis of critical topics such as misinformation spread, audience engagement, and coordinated messaging, while also capturing hashtag co-occurrence networks and trends over time. By making this dataset publicly available, we aim to empower researchers to explore TikTok's role in shaping electoral discourse and its broader impact on democracy and public opinion.

\subsection*{Limitations}
The TikTok Research API imposes constraints such as rate limits, token expiration, and daily request caps, which made large-scale data collection time-intensive. To mitigate these challenges, we prioritized video metadata collection and expanded our team to distribute the workload, ensuring more efficient data retrieval. 



\subsection*{Future Work}
Data collection will conclude a week after Inauguration Day in January 2025. To analyze the online discourse surrounding the election and its aftermath, we will prioritize collecting comments and replies from recent months (starting in May) and work backward as the process becomes more efficient. Our public repository will be continuously updated with video IDs, detected transcript languages, comment summaries, and AI-generated video descriptions. This evolving dataset will enable meaningful analysis of public sentiment and discussion trends related to the election.

\subsection*{Data Availability} 
The dataset, available via GitHub, complies with TikTok's Terms and Conditions, which restrict the direct distribution of collected videos. To adhere to these guidelines, we provide Video IDs—unique identifiers for each post—rather than the videos themselves. These IDs enable researchers to use TikTok’s API to retrieve complete video objects, including associated multimedia and metadata. 
The repository, that is available at \href{https://github.com/gabbypinto/US2024PresElectionTikToks}{https://github.com/gabbypinto/US \\2024PresElectionTikToks}, will be continuously updated to ensure the dataset remains current and accessible.



\bibliographystyle{ACM-Reference-Format}
\bibliography{sample-base}



\end{document}